\documentclass[aps,prl,floatfix,twocolumn]{revtex4}
\usepackage{amsmath}
\usepackage{graphicx}
\usepackage{epstopdf}

\begin{document}
\title{Manipulating biphotonic qutrits}
\author{B. P. Lanyon, T. J. Weinhold, N. K. Langford, J. L. O'Brien$^\dagger$, K. J. Resch$^\ddagger$, A. Gilchrist,  and A. G. White}
\affiliation{Department of Physics and Centre for Quantum Computer Technology, University of Queensland, QLD 4072, Brisbane, Australia.\\
$\dagger$Centre for Quantum Photonics, H. H. Wills Physics Laboratory \& Department of Electrical and Electronic Engineering, University of Bristol, Merchant Venturers Building, Woodland Road, Bristol, BS8 1UB, UK\\
$\ddagger$ Institute for Quantum Computing and Department of Physics \& Astronomy, University of Waterloo, Waterloo, N2L 3G1, Canada.}

\begin{abstract}
\noindent Quantum information carriers with higher dimension than the canonical qubit offer significant advantages. However, manipulating such systems is extremely difficult. We show how measurement induced non-linearities can be employed to dramatically extend the range of possible transforms on 
biphotonic qutrits; the three level quantum systems formed by the polarisation of two photons in the same spatio-temporal mode. We fully characterise the biphoton-photon entanglement that underpins our technique, thereby realising the first instance of qubit-qutrit entanglement. We discuss an extension of our technique to generate qutrit-qutrit entanglement and to manipulate any bosonic encoding of quantum information.
\end{abstract}
\maketitle

Higher dimensional systems offer advantages such as increased-security in a range of quantum information protocols  \cite{langford:053601, PhysRevA.65.012310, molina-terriza:167903, 1367-2630-8-5-075, PhysRevLett.88.127901, PhysRevLett.88.127902, PhysRevA.67.012311}, greater channel capacity for quantum communication \cite{PhysRevLett.90.167906}, novel fundamental tests of quantum mechanics \cite{PhysRevLett.88.040404, PhysRevA.65.032118} and more efficient quantum gates \cite{ralph:022313}. Optically such systems have been realised using polarisation \cite{PhysRevLett.47.460} and transverse spatial modes \cite{Mair:2001fk, langford:053601}. However in each case state transformation techniques have proved difficult to realise. In fact, performing such transformations is a significant problem in a range of physical architectures.

The polarisation of two photons in the same spatio-temporal mode represents a three-level bosonic quantum system, a biphotonic qutrit, with symmetric logical basis states: $|0_3\rangle{\equiv} |2_H,0_V\rangle$, $|1_3\rangle{\equiv}(|1_H,1_V\rangle{+}|1_V,1_H\rangle)/\sqrt{2}$, and $|2_3\rangle{\equiv} |0_H,2_V\rangle$ \cite{2005SPIE.5833..202B}.  The simple optical tools which allow full control over the polarisation of a photonic qubit are insufficient for full control over a biphotonic qutrit \cite{bogdanov:042303}.  Consequently even simple state transformations required in qutrit generation, processing and measurement  are extremely limited. Significant progress has been made in biphoton state generation. For example, complex arbitrary state preparation techniques that employ multiple nonlinear crystals \cite{bogdanov:230503}  and non-maximally entangled states \cite{dariano.062337}  have been developed.

Here we present and demonstrate a technique that  dramatically extends the range of biphotonic qutrit transforms, for use in all stages of qutrit manipulation. The technique is based on a Fock-state filter which employs a measurement-induced nonlinearity to conditionally remove photon-number (Fock) states from superpositions \cite{PhysRevA.66.064303, PhysRevLett.88.147901, PhysRevA.66.064302, sanaka:017902, sanaka:083601, resch:203602}. We first demonstrate the action of the filter as a qutrit polariser, which can conditionally remove a single logical qutrit state from a superposition. We then combine this nonlinear operation with standard waveplate rotations to demonstrate the dramatically increased range of qutrit transforms it enables. 
Finally we present the first instance and full characterisation of a polarisation entangled photon-biphoton state, which underpins the power of our technique. Such qubit-qutrit states have been studied extensively \cite{PhysRevLett.77.1413, slater-2003-5, cabello-2005-72, Jami:quant-ph0606039, osenda-2005, slater-2007, ann-2007} and we suggest an extension to generate qutrit-qutrit entanglement.

\begin{figure}%[!t]
\vspace{-5mm}
\includegraphics[width=1\columnwidth]{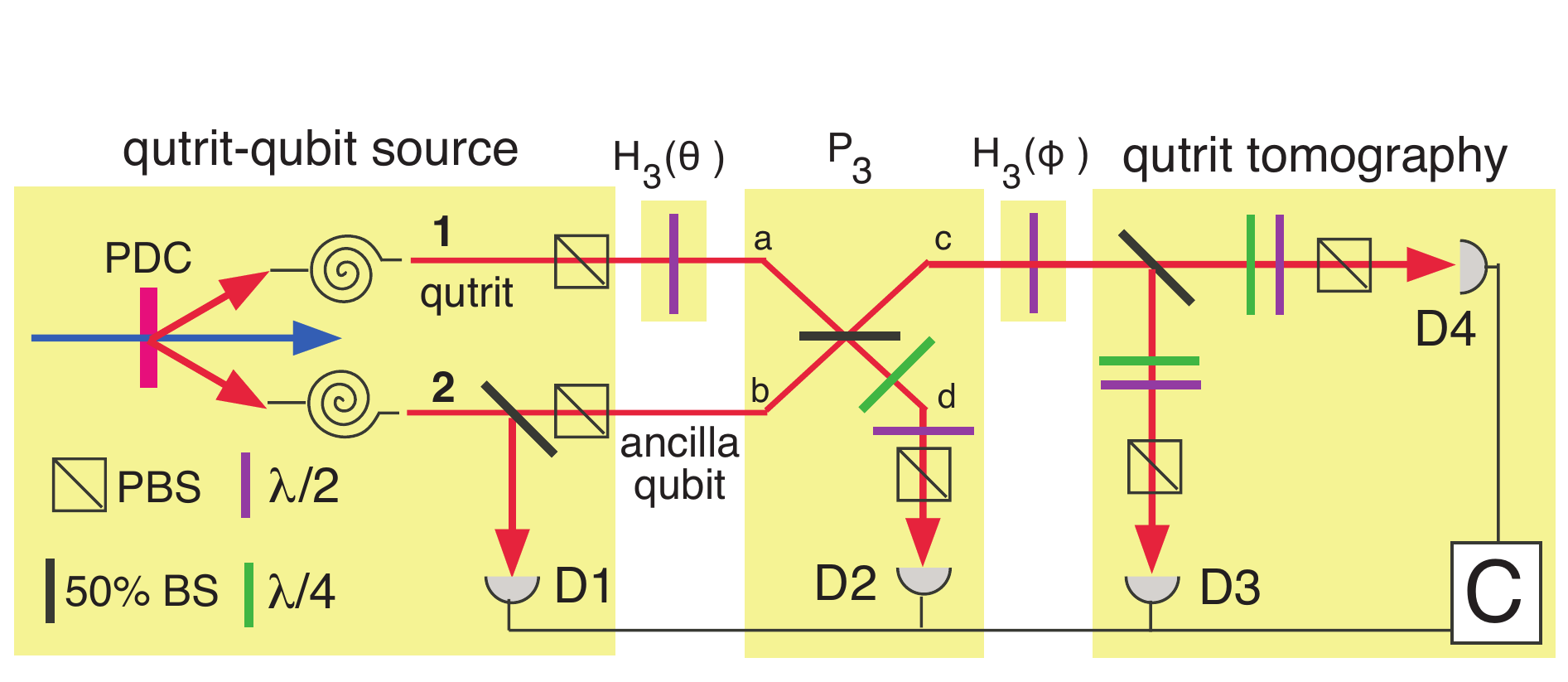} \vspace{-5mm}
\caption{Experimental schematic. Emission from a parametric down conversion (PDC) crystal is coupled into single mode fibre and injected into modes 1 and 2. Coincident (C) detection of photons at D1-4 selects, with high probability,  the cases of double photon-pair emission from the PDC source.
\vspace{-5mm}
}
\label{fig1:algo}
\end{figure}

We generate our qutrits through double-pair emission from a spontaneous parametric down conversion source (Fig.~1). Measurement of a four-fold coincidence between detectors D1-D4 selects, with high probability, the cases of double-pair emission into inputs 1 and 2. The biphoton state in mode 1 is passed through a horizontal polariser to prepare the logical qutrit state $|0_3\rangle$.
Input 2 is passed through a 50\% beam splitter; detection of a single photon at D1 signals, with high probability, a single photon in mode $b$; which is passed through a polarising beam splitter to prepare a polarisation ancilla qubit ($|0_2\rangle{\equiv} |1_{H} \rangle$, $|1_2\rangle{\equiv} |1_{V} \rangle$) in the logical state $|0_2\rangle$. Thus a qubit and qutrit arrive simultaneously at the central 50\% beam splitter.

The operation of a Fock-filter relies on nonclassical interference effects \cite{PhysRevLett.59.2044}. When two indistinguishable photons are injected into modes $a$ and $b$ (Fig.~1), the probability of detecting a single photon in mode $d$ is zero; if two or more photons are injected into mode $a$ then this probability is non-zero. By injecting a single photon into mode $b$ and detecting a single photon in mode $d$, single photon terms can therefore be removed from any photon number superposition states arriving in mode $a$. In fact, by varying the reflectivity of the beam splitter it is possible to conditionally remove any number-state from a superposition \cite{sanaka:083601}. This Fock-state filter acts only on light with the same polarisation as the ancilla (in our case, horizontal), so by detecting a single horizontal photon in mode $d$, the logical qutrit state $|1_3\rangle$ is blocked, since it contains a single photon with the same polarisation as the ancilla. The remaining logical qutrit states are only attenuated. 

Thus for a beam splitter of reflectivity $50\%$ the filter acts, on an initial qutrit state of $\alpha|0_3\rangle{+}\beta|1_3\rangle{+}\gamma|2_3\rangle$, as a qutrit polariser described by the operator $\mathbf{P}_3{=}|0_3\rangle\langle 0_3| {-} |2_3\rangle\langle 2_3|$. Note that a standard horizontal or vertical linear polariser would act as a lossless qutrit polariser, but one restricted to simultaneously removing either the logical $| 0_3 \rangle$ and $|1_3\rangle$, or the $|1_3\rangle$ and $|2_3\rangle$ states. Such a
polariser could therefore not leave just the $|1_3\rangle$ state. By varying the polarisation of the ancilla, and the reflectivity of the central beam splitter, the operation of our lossy qutrit polariser can be tuned to preferentially remove the $| 0_3 \rangle$, $| 1_3 \rangle$ or $| 2_3 \rangle$ states. We choose to demonstrate removal of the $| 1_3 \rangle$ state and include the general operation of the filter for an arbitrary beam splitter reflectivity \cite{Fockfilter}.

The qutrit polariser offers a powerful tool for transforming between qutrit states. For example, consider the initial qutrit state  $|0_3\rangle$ injected into input 1, the red dot of Fig.~2a). The black ring shows the limited range of qutrit states, with real coefficients, that are accessible using waveplates \cite{waveplates}. By including the qutrit polariser the range is dramatically extended to the closed sphere in Fig~2; the transformation to any real state is possible.

We measure our qutrits by  passing mode $c$ through a 50\% beam splitter and performing polarisation analysis of the two outputs in coincidence, as shown in Fig.~1. This analysis non-deterministically discriminates the logical states $|0_3\rangle$, $|1_3\rangle$, and $|2_3\rangle$ with probabilities $p(0_3){=}\frac{1}{2}$, $p(1_3){=}\frac{1}{4}$ and $p(2_3){=}\frac{1}{2}$. Combining it with single qubit rotations after the beam splitter allows us to perform full qutrit state tomography of mode $c$. Complete qutrit tomography requires nine independent measurements, which we construct from logical basis states and two-part superpositions \cite{langford:053601}. Our method differs from that of Refs \cite{2005SPIE.5833..202B, bogdanov:042303} and is described in additional online material.  We use convex optimisation to reconstruct the qutrit density matrix and Monte-Carlo simulations for error analysis \cite{obrien:080502, Gilchrist:2007lr}.

\begin{figure}%[!t]
\includegraphics[width=0.6\columnwidth]{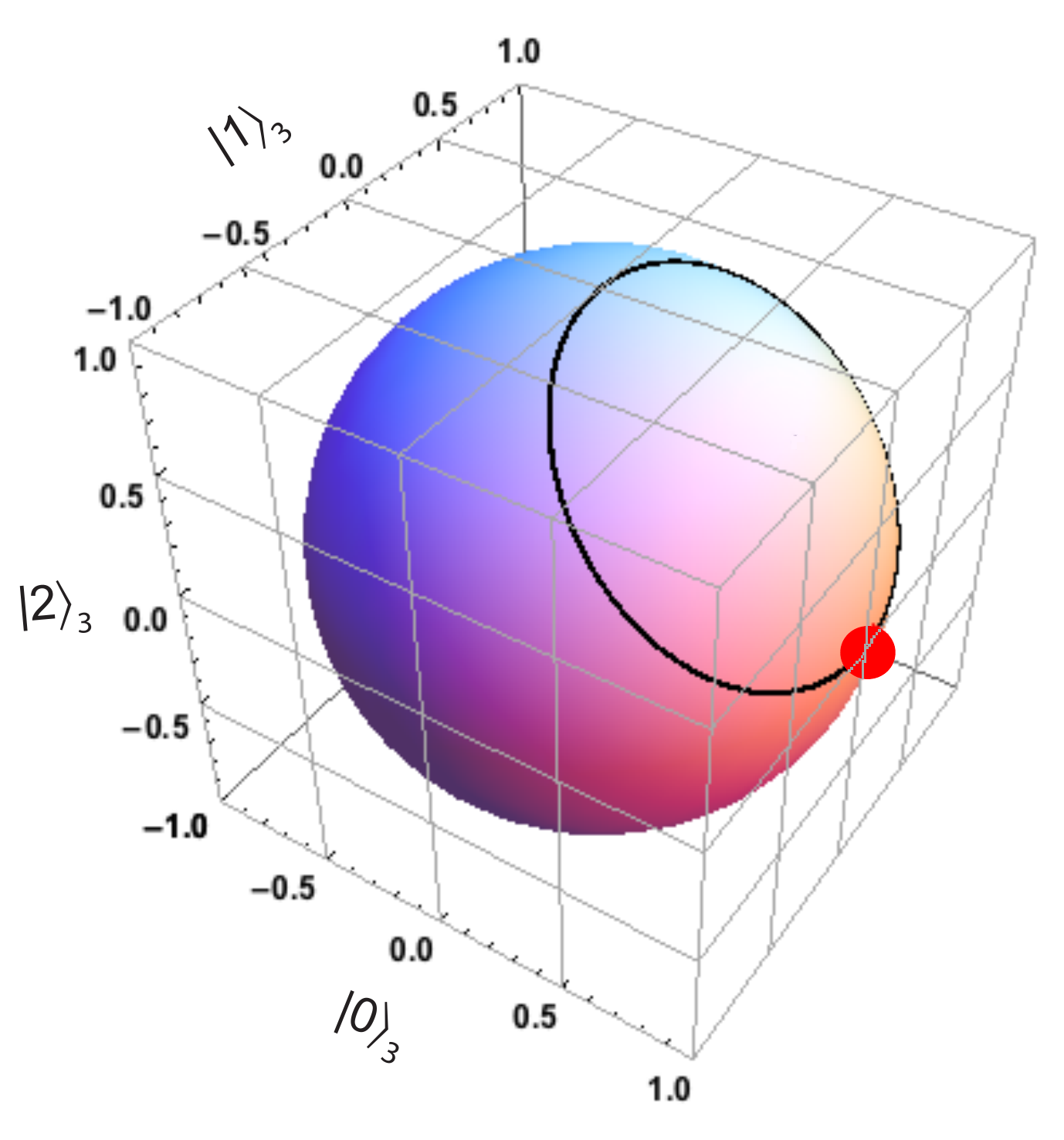} \vspace{-5mm}
\caption{Comparison of the range of linearly polarised qutrit states achievable by transforming the state $| 0_3 \rangle$ (red dot); when using only waveplate operations (black ring);  by incorporating our qutrit polariser,$\mathbf{Q}_{3}(\alpha)\mathbf{H}_{3}(\phi) \mathbf{P}_{3}(\sqrt{0.5}) \mathbf{H}_{3}(\theta)|0_3\rangle$ (sphere) \cite{Fockfilter, waveplates}. }
\label{fig2:algo}
\vspace{-5mm}
\end{figure}

Ideally both the central and tomography beam splitters reflect 50\% of both polarisations.  In practice, we found that they deviated by a few percent and impart undesired unitary rotations on the reflected and transmitted modes. For the tomography beam splitter, these imperfections modified the nine measured qutrit states; we characterised this effect and incorporated it into the tomographic reconstruction. We also found that the effect of the imperfect central beam splitter on the performance of the qutrit polariser was negligible.

A frequency-doubled mode-locked Ti:Sapphire laser (820 nm$\rightarrow$410 nm, $\Delta \tau{=}80$fs at 82 MHz repetition rate) is used to produce photon pairs via parametric down conversion from a Type I phase-matched 2mm Bismuth Borate (BiBO) crystal, filtered by blocked interference filters (820$\pm$1.5 nm). We collect the down-conversion into single-mode optical fibers. Photons are detected using fiber-coupled single photon counting modules and coincidences measured using a Labview (National Instruments) interfaced quad-logic card (ORTEC CO4020). When directly coupled into detectors the source yielded two-folds at 60 kHz and singles rates at 220 kHz. At the output of the complete circuit we observed four-fold coincidence rates at approximately 1 Hz.

The quality of the non-classical interference underpinning the qutrit polariser can be measured directly \cite{sanaka:083601}. Ref.~\cite{resch:203602} relates non-classical visibilities to a Fock-state filter's ability to block single photon terms. Using this technique, and measured visibilities of $V_{1}{=}97\pm{1}\%$ \& $V_{2}{=}68\pm{4}\%$, we predict an extinction ratio of $5(\pm{2}){:}1$; our qutrit polariser will pass the logical $|0_3\rangle$ and $|2_3\rangle$ states at five times the rate it passes the logical $|1_3\rangle$ state.

To demonstrate the qutrit polariser we include a half-waveplate in mode $a$ set to $\theta{=}\frac{\pi}{8}$ to generate a superposition qutrit state with all logical states populated, of the form \cite{waveplates}:
\begin{equation}
\mathbf{H}_{3}(\theta)|0_3\rangle{=}\cos^2 2\theta |0_3\rangle{+}\sin^2 2\theta |2_3\rangle {+} \sin 4\theta |1_3\rangle/\sqrt{2}
\end{equation}

\noindent We measure the output state in mode $c$ without applying the qutrit polariser. This is achieved by blocking the ancilla photon in mode $b$, and performing qutrit tomography of mode $c$ in two-fold coincidence between D3 and D4. The experimentally reconstructed density matrix is shown in Fig.~3a) and has a near perfect fidelity between the measured and ideal states, $F{=} 97{\pm}1\%$, and a low linear entropy, $S_L{=} 6{\pm}7\%$, \cite{definitions, measuringgates}. We then prepared the output state by unblocking the ancilla and, as in all further cases, perform tomography of mode $c$ in four-fold coincidence between D1-4. The qutrit polariser is now `on' and we expect the absorption of the logical $|1_3\rangle$ state. 
The reconstructed density matrix is shown in Fig.~3b) and has a lower fidelity with the ideal,  $F{=} 78{\pm}8\%$, and linear entropy $S_L{=} 47{\pm}14\%$. The relative reduction in the logical $|1_{3}\rangle$ state probability, when the filter is turned on, yields an extinction ratio of $6.80(\pm{0.07}){:}1$, consistent with that predicted above.
\begin{figure}
\includegraphics[width=1\columnwidth]{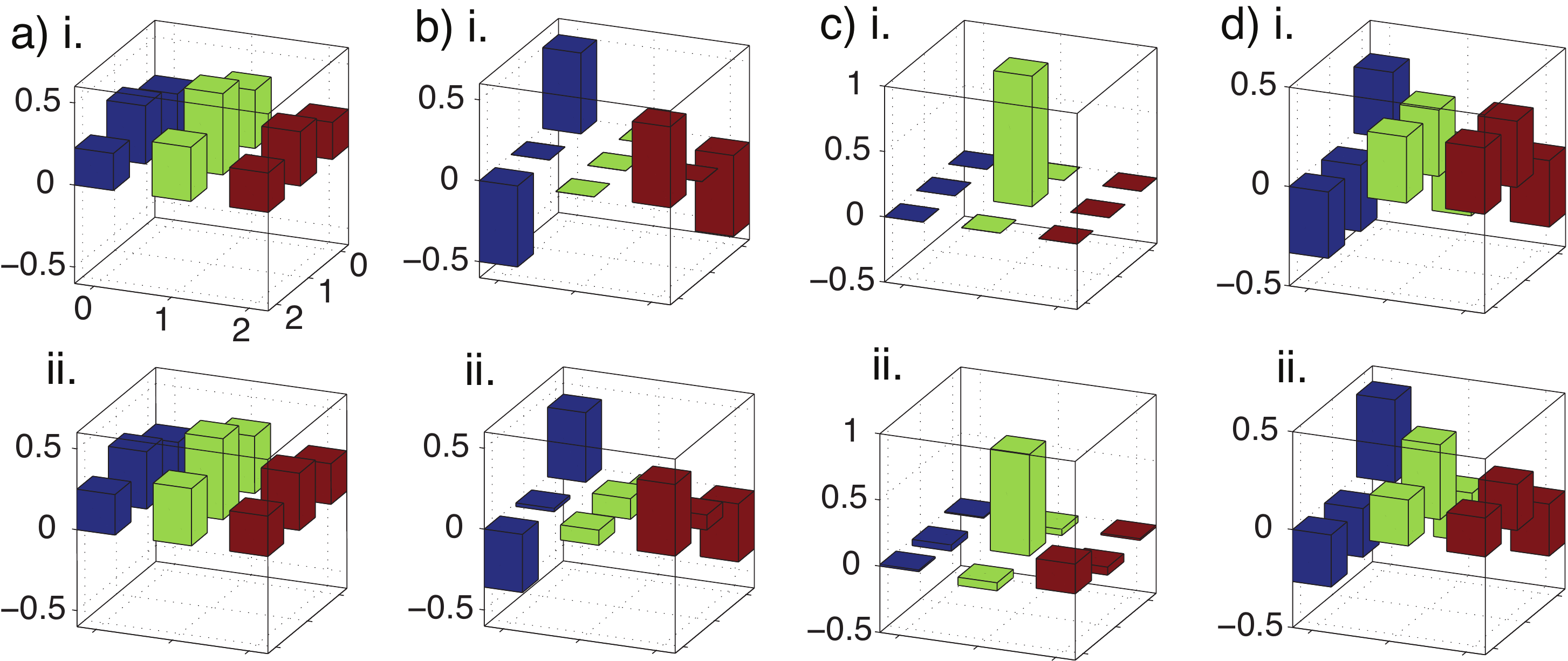}
\vspace{-5mm}
\caption{Comparison of real parts of (i) ideal and (ii) measured qutrit density matrices.  
a) The measured output state with the qutrit polariser `off' (Eq.~1 for $\theta{=}\frac{\pi}{8}$). b) The output state with the qutrit polariser `on' showing the removal of the logical $|1_3\rangle$ qutrit state. c)-d) Newly accessible qutrit states $|1_3\rangle$ and ($|0_3\rangle{-}|1_3\rangle{-}|2_3\rangle$)/$\sqrt{3}$, respectively. States b-d) all lie on the surface of the sphere of Fig.~2, but not on the ring.
\vspace{-5mm}
}
\label{fig3:newstates}
\end{figure}

Measured non-classical visibilities are significantly limited by higher-order parametric down conversion photon number terms \cite{bristolpaper, Weinhold}. After removing these effects, as described in reference \cite{resch:203602}, we find a corrected two-fold visibility of $V'_1{=}100\pm{1}\%$, that would be measured given an ideal two-photon source (higher-order effects cannot be distinguished from experimental uncertainty in the four-fold visibility). This corrected visibility can then be used to predict the potential performance of our circuit given an ideal photon source \cite{resch:203602}; in this case we predict that the filter would pass the logical $|0_3\rangle$ and $|2_3\rangle$ states at least $24$ times the rate it passes the logical $|1_3\rangle$ state. Clearly the performance of our qutrit polariser is significantly limited by higher-order emissions from our optical source.%%

Figures~3c-d) show experimentally reconstructed density matrices of newly accessible states achieved by incorporating the qutrit polariser with half-waveplate operations applied to the initial state of $|0_3\rangle$; $|1_3\rangle$ and ($|0_3\rangle{-}|1_3\rangle{-}|2_3\rangle$)/$\sqrt{3}$. The fidelities with the ideal are $77{\pm}3\%$, and $83{\pm}7\%$ with linear entropies  $51{\pm}7\%$ and $38{\pm}15\%$, respectively. These fidelities exceed the maximum achievable using only linear waveplates (50\%) by $9{\pm}1$ and $5{\pm}1$ standard deviations, respectively.

The qutrit polariser employs a measurement-induced non-linearity whereby the biphoton becomes entangled with the ancilla photon. Instead of detecting the ancilla in a single, fixed polarisation state, we can also use tomographic measurements to directly investigate this resultant \emph{entangled qubit-qutrit system}. Without emphasis to the physical systems involved, such states where first studied by Peres as a special case of his negativity criterion for entanglement; a negativity of 0 ($>$0) is conclusive of a separable (entangled) state \cite{negativity, PhysRevLett.77.1413}.  More recently these states have received a significant amount of attention \cite{PhysRevLett.77.1413, slater-2003-5, cabello-2005-72, Jami:quant-ph0606039, osenda-2005, slater-2007}  and have been predicted to exhibit novel entanglement sudden death phenomena \cite{ann-2007} .

On injection of the qutrit state given by Eq.~1 into the Fock-filter, we find the following qubit-qutrit joint state of modes $c$ and $d$:
\begin{eqnarray}
\{ \cos^{2} 2\theta |0_{2}, 0_{3} \rangle  +\sin 4\theta |1_{2}, 0_{3} \rangle & \\
+ \sin^{2} 2\theta (\sqrt{2}|1_{2}, 1_{3} \rangle - |0_{2}, 2_{3} \rangle) \} / N, & \nonumber
\end{eqnarray}

\noindent where $N{=} \sqrt{2 {-} \cos 4 \theta}$. By varying $\theta$ we can tune the level of entanglement from zero ($\theta{=}0$) to near-maximal ($\theta{=}\frac{\pi}{4}$), with corresponding negativities of $0$ to $\sqrt{8/9}\approx 0.94$, respectively. To perform qubit-qutrit state tomography we use 36 independent measurements constructed from all of the combinations of the aforementioned nine qutrit states and four qubit states [H,V,D,R]. Fig.~4 shows the measured density matrix for the near-maximally entangled case, which corresponds to the preparation of two vertically polarised photons in mode $a$. There is a high fidelity with the ideal of $81{\pm}3\%$, low linear entropy of $17{\pm}5\%$ and the state is highly entangled with a negativity of  $0.77{\pm}0.05$. We note that a maximally entangled state is predicted for $\theta{=}\frac{\pi}{4}$ and a central beam splitter reflectivity of $R{=}\sqrt{2}/(\sqrt{2}{+}1)\approx58.6\%$.

Entangling information carriers to ancilla qubits is an extremely powerful technique \cite{Knill:2001qy}: such correlations play a central role in the power of the Fock filter to  transform biphotonic qutrits. However, the application of our technique is not limited to extending transforms on single qutrits. We propose that the generation of qubit-qutrit entanglement offers a path to realise multi-qutrit operations. For example, a pair of entangled qubit-qutrit states could be used to create qutrit-qutrit entanglement by projecting the qubits into an entangled state using well-known techniques. The much anticipated development of high-brightness single-photon sources will make such experiments feasible in the near future. We wish to emphasize that our technique is not limited to manipulating biphotons. The Fock-filter can be applied to any system where measurement can induce non-linear effects; that is any bosonic encoding of quantum information, including bosonic atoms \cite{popescu-2006} and time-bin, frequency and orbital angular momentum encoding of photons.

\begin{figure}
\includegraphics[width=\columnwidth]{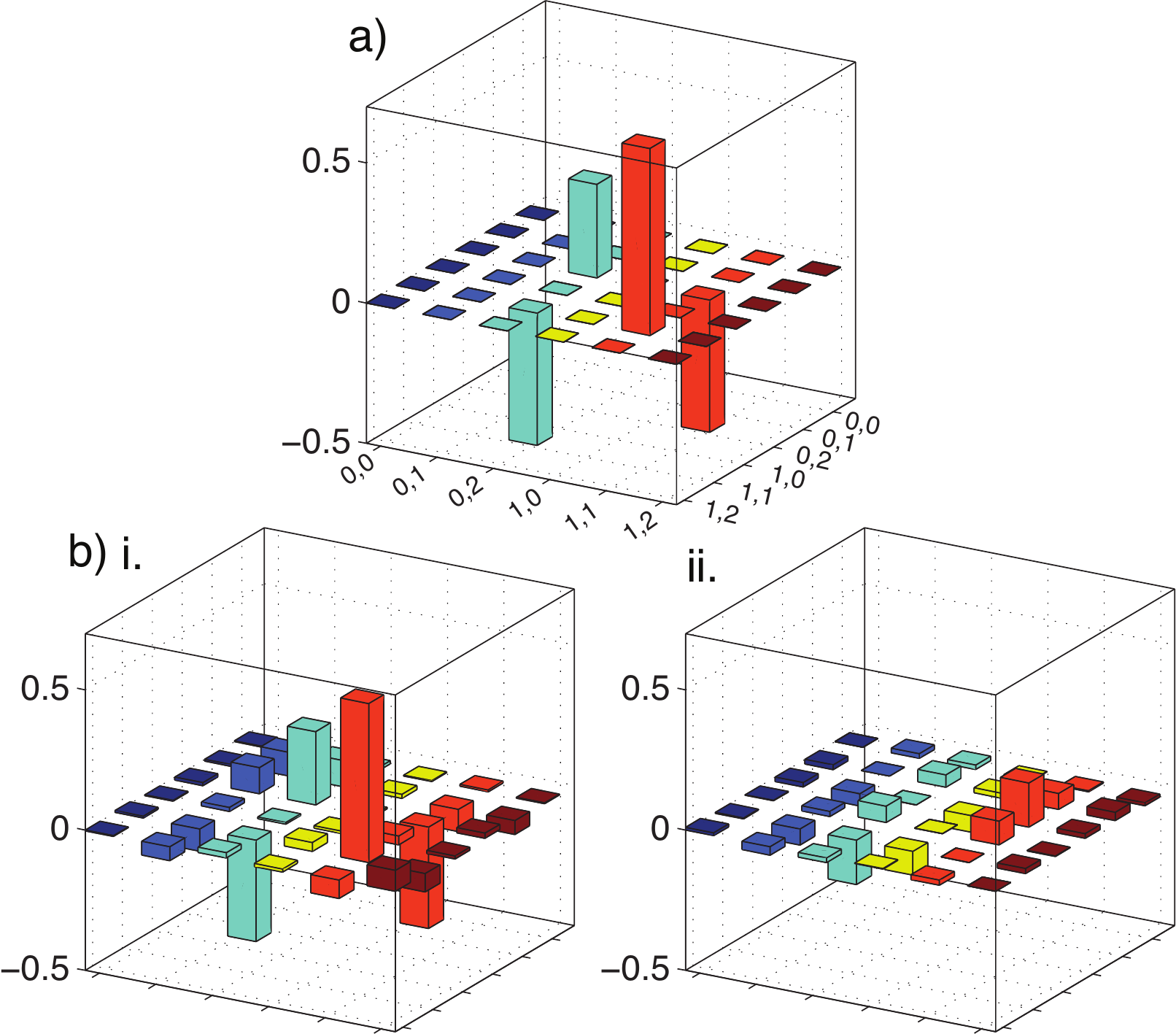}\
\vspace{-6mm}
\caption{Comparison of entangled qubit-qutrit density matrices. a) ideal, b) \& c) measured real \&  imaginary parts. There is high fidelity with the ideal ($81 \pm{3}\%$), low linear entropy ($17\pm{5}\%$) and the state is highly entangled with a negativity of $0.77\pm{0.05}$. The ideal state is given by Eq.~2 for $\theta{=}\pi/4$. Note the axis label: $x, j$ represents the qubit logical state $x$ and the qutrit logical state $j$ i.e. $|x_2, j_3\rangle$.
\vspace{-6mm}
}
\label{fig3:qubit-qutrit}
\end{figure}

We have shown that measurement induced nonlinearities offer significant advantages for the manipulation of higher dimensional bosonic information carriers, specifically biphotonic qutrits. We demonstrated a nonlinear qutrit-polariser, capable of conditionally removing a single logical qutrit state from a superposition, and greatly extending the range of possible qutrit transforms. Such tools could find immediate application to quickly generate the mutually unbiased basis states required for optimum security in quantum key distribution protocols involving qutrits \cite{PhysRevLett.88.127901, PhysRevLett.88.127902, PhysRevA.67.012311} or as a filtering technique to manipulate entanglement in qutrit-qutrit states. 
Finally we fully characterised the entangled photon-biphoton state that underpins the power of our technique. This is the first instance of the generation and characterisation of entanglement between these distinct physical systems and makes recent theoretical proposals experimentally testable \cite{ann-2007}. Besides offering a path to implement novel multi-qutrit operations we propose that our technique can be extended to manipulate any bosonic encoding of quantum information. 

NOTE: After completion of this work , several authors presented proposals for which our technique 
is directly relevant \cite{bregman-2007,bishop2007,Ali07}.

This work was supported in part by the Australian Research Office and the US Disruptive Technologies Office. BPL and AGW wish to acknowledge funding by the Endeavor Europe Award and Federation Fellow programs, respectively.\\


\vspace{-1cm}

\footnotesize
\begin{thebibliography}{34}
\expandafter\ifx\csname natexlab\endcsname\relax\def\natexlab#1{#1}\fi
\expandafter\ifx\csname bibnamefont\endcsname\relax
  \def\bibnamefont#1{#1}\fi
\expandafter\ifx\csname bibfnamefont\endcsname\relax
  \def\bibfnamefont#1{#1}\fi
\expandafter\ifx\csname citenamefont\endcsname\relax
  \def\citenamefont#1{#1}\fi
\expandafter\ifx\csname url\endcsname\relax
  \def\url#1{\texttt{#1}}\fi
\expandafter\ifx\csname urlprefix\endcsname\relax\def\urlprefix{URL }\fi
\providecommand{\bibinfo}[2]{#2}
\providecommand{\eprint}[2][]{\url{#2}}

 \bibitem[{\citenamefont{Langford et~al.}(2004)\citenamefont{Langford, Dalton,
  Harvey, O'Brien, Pryde, Gilchrist, Bartlett, and White}}]{langford:053601}
\bibinfo{author}{\bibfnamefont{N.~K.} \bibnamefont{Langford \emph{et al}}},
  \bibinfo{journal}{Phys. Rev. Lett.} \textbf{\bibinfo{volume}{93}},
  \bibinfo{eid}{053601} (\bibinfo{year}{2004}).
 
 \bibitem[{\citenamefont{Spekkens and Rudolph}(2001)}]{PhysRevA.65.012310}
\bibinfo{author}{\bibfnamefont{R.~W.} \bibnamefont{Spekkens}} \bibnamefont{and}
  \bibinfo{author}{\bibfnamefont{T.}~\bibnamefont{Rudolph}},
  \bibinfo{journal}{Phys. Rev. A} \textbf{\bibinfo{volume}{65}},
  \bibinfo{pages}{012310} (\bibinfo{year}{2001}).

\bibitem[{\citenamefont{Molina-Terriza
  et~al.}(2004)\citenamefont{Molina-Terriza, Vaziri, Rehacek, Hradil, and
  Zeilinger}}]{molina-terriza:167903}
\bibinfo{author}{\bibfnamefont{G.}~\bibnamefont{Molina-Terriza \emph{et al}}},
  \bibinfo{journal}{Phys. Rev. Lett.} \textbf{\bibinfo{volume}{92}},
  \bibinfo{eid}{167903} (\bibinfo{year}{2004}).

\bibitem[{\citenamefont{Gr\"{o}blacher
  et~al.}(2006)\citenamefont{Gr\"{o}blacher, Jennewein, Vaziri, Weihs, and
  Zeilinger}}]{1367-2630-8-5-075}
\bibinfo{author}{\bibfnamefont{S.}~\bibnamefont{Gr\"{o}blacher \emph{et al}}},
  \bibinfo{journal}{New J. Phys.} \textbf{\bibinfo{volume}{8}},
  \bibinfo{pages}{75} (\bibinfo{year}{2006}).

\bibitem[{\citenamefont{Bru\ss{} and
  Macchiavello}(2002)}]{PhysRevLett.88.127901}
\bibinfo{author}{\bibfnamefont{D.}~\bibnamefont{Bru\ss{}}} \bibnamefont{and}
  \bibinfo{author}{\bibfnamefont{C.}~\bibnamefont{Macchiavello}},
  \bibinfo{journal}{Phys. Rev. Lett.} \textbf{\bibinfo{volume}{88}},
  \bibinfo{pages}{127901} (\bibinfo{year}{2002}).

\bibitem[{\citenamefont{Cerf et~al.}(2002)\citenamefont{Cerf, Bourennane,
  Karlsson, and Gisin}}]{PhysRevLett.88.127902}
\bibinfo{author}{\bibfnamefont{N.~J.} \bibnamefont{Cerf \emph{et al}}},
  \bibinfo{journal}{Phys. Rev. Lett.} \textbf{\bibinfo{volume}{88}},
  \bibinfo{pages}{127902} (\bibinfo{year}{2002}).

\bibitem[{\citenamefont{Durt et~al.}(2003)\citenamefont{Durt, Cerf, Gisin, and
  \ifmmode~\dot{Z}\else \.{Z}\fi{}ukowski}}]{PhysRevA.67.012311}
\bibinfo{author}{\bibfnamefont{T.}~\bibnamefont{Durt \emph{et al}}},
  \bibinfo{journal}{Phys. Rev. A}
  \textbf{\bibinfo{volume}{67}}, \bibinfo{pages}{012311}
  (\bibinfo{year}{2003}).

\bibitem[{\citenamefont{Fujiwara et~al.}(2003)\citenamefont{Fujiwara, Takeoka,
  Mizuno, and Sasaki}}]{PhysRevLett.90.167906}
\bibinfo{author}{\bibfnamefont{M.}~\bibnamefont{Fujiwara \emph{et al}}},
  \bibinfo{journal}{Phys. Rev. Lett.} \textbf{\bibinfo{volume}{90}},
  \bibinfo{pages}{167906} (\bibinfo{year}{2003}).

\bibitem[{\citenamefont{Collins et~al.}(2002)\citenamefont{Collins, Gisin,
  Linden, Massar, and Popescu}}]{PhysRevLett.88.040404}
\bibinfo{author}{\bibfnamefont{D.}~\bibnamefont{Collins \emph{et al}}},
  \bibinfo{journal}{Phys. Rev. Lett.} \textbf{\bibinfo{volume}{88}},
  \bibinfo{pages}{040404} (\bibinfo{year}{2002}).

\bibitem[{\citenamefont{Kaszlikowski et~al.}(2002)\citenamefont{Kaszlikowski,
  Kwek, Chen, \ifmmode~\dot{Z}\else \.{Z}\fi{}ukowski, and
  Oh}}]{PhysRevA.65.032118}
\bibinfo{author}{\bibfnamefont{D.}~\bibnamefont{Kaszlikowski \emph{et al}}},
  \bibinfo{journal}{Phys. Rev. A}
  \textbf{\bibinfo{volume}{65}}, \bibinfo{pages}{032118},
  (\bibinfo{year}{2002}).

\bibitem[{\citenamefont{Ralph et~al.}(2007)\citenamefont{Ralph, Resch, and
  Gilchrist}}]{ralph:022313}
\bibinfo{author}{\bibfnamefont{T.~C.} \bibnamefont{Ralph \emph{et al}}},
  \bibinfo{journal}{Phys Rev A}
  \textbf{\bibinfo{volume}{75}}, \bibinfo{eid}{022313}
    (\bibinfo{year}{2007}).

  \bibitem[{\citenamefont{Aspect et~al.}(1981)\citenamefont{Aspect, Grangier, and
  Roger}}]{PhysRevLett.47.460}
\bibinfo{author}{\bibfnamefont{A.}~\bibnamefont{Aspect \emph{et al}}},
  \bibinfo{journal}{Phys. Rev. Lett.} \textbf{\bibinfo{volume}{47}},
  \bibinfo{pages}{460} (\bibinfo{year}{1981}).

\bibitem[{\citenamefont{Mair et~al.}(2001)\citenamefont{Mair, Vaziri, Weihs,
  and Zeilinger}}]{Mair:2001fk}
\bibinfo{author}{\bibfnamefont{A.}~\bibnamefont{Mair \emph{et al}}},
  \bibinfo{journal}{Nature} \textbf{\bibinfo{volume}{412}},
  \bibinfo{pages}{313} (\bibinfo{year}{2001}).

\bibitem[{\citenamefont{{} et~al.}(2005)\citenamefont{{Bogdanov},
  {Chekhova}, {Kulik}, {Maslennikov}, {Oh}, and {Tey}}}]{2005SPIE.5833..202B}
\bibinfo{author}{\bibfnamefont{Y.}~\bibnamefont{{Bogdanov \emph{et al}}}},
  \emph{\bibinfo{booktitle}{Quantum Informatics 2004}},
 pp. \bibinfo{pages}{202--212},
   (\bibinfo{year}{2005}).

 \bibitem[{\citenamefont{Bogdanov
  et~al.}(2004{\natexlab{a}})\citenamefont{Bogdanov, Chekhova, Krivitsky,
  Kulik, Penin, Zhukov, Kwek, Oh, and Tey}}]{bogdanov:042303}
\bibinfo{author}{\bibfnamefont{Y.~I.} \bibnamefont{Bogdanov}},
  \bibinfo{journal}{Phys. Rev. A},
  \textbf{\bibinfo{volume}{70}}, \bibinfo{eid}{042303},
 \bibinfo{year}{(2004)}.

\bibitem[{\citenamefont{Bogdanov
  et~al.}(2004{\natexlab{b}})\citenamefont{Bogdanov \emph{et al}}}]{bogdanov:230503}
\bibinfo{author}{\bibfnamefont{Y.~I.} \bibnamefont{Bogdanov}},
  \bibinfo{journal}{Phy. Rev. Lett.} \textbf{\bibinfo{volume}{93}},
\bibinfo{year}{(2004)}.

\bibitem[{\citenamefont{Vallone}(2005)\citenamefont{Vallone}}]{dariano.062337}
\bibinfo{author}{\bibfnamefont{G.} \bibnamefont{Vallone \emph{et al}}},
  \bibinfo{journal}{Phys. Rev. A} \textbf{\bibinfo{volume}{76}},
  \bibinfo{eid}{012319} (\bibinfo{year}{2007}).

\bibitem[{\citenamefont{Grudka and W\'ojcik}(2002)}]{PhysRevA.66.064303}
\bibinfo{author}{\bibfnamefont{A.}~\bibnamefont{Grudka}} \bibnamefont{and}
  \bibinfo{author}{\bibfnamefont{A.}~\bibnamefont{W\'ojcik}},
  \bibinfo{journal}{Phys. Rev. A} \textbf{\bibinfo{volume}{66}},
  \bibinfo{pages}{064303} (\bibinfo{year}{2002}).

\bibitem[{\citenamefont{Hofmann and Takeuchi}(2002)}]{PhysRevLett.88.147901}
\bibinfo{author}{\bibfnamefont{H.~F.} \bibnamefont{Hofmann}} \bibnamefont{and}
  \bibinfo{author}{\bibfnamefont{S.}~\bibnamefont{Takeuchi}},
  \bibinfo{journal}{Phys. Rev. Lett.} \textbf{\bibinfo{volume}{88}},
  \bibinfo{pages}{147901} (\bibinfo{year}{2002}).

\bibitem[{\citenamefont{Zou et~al.}(2002)\citenamefont{Zou, Pahlke, and
  Mathis}}]{PhysRevA.66.064302}
\bibinfo{author}{\bibfnamefont{X.}~\bibnamefont{Zou \emph{et al}}},
  \bibinfo{journal}{Phys. Rev. A} \textbf{\bibinfo{volume}{66}},
  \bibinfo{pages}{064302} (\bibinfo{year}{2002}).

\bibitem[{\citenamefont{Sanaka et~al.}(2004)\citenamefont{Sanaka, Jennewein,
  Pan, Resch, and Zeilinger}}]{sanaka:017902}
\bibinfo{author}{\bibfnamefont{K.}~\bibnamefont{Sanaka \emph{et al}}},
  \bibinfo{journal}{Phys. Rev. Lett.} \textbf{\bibinfo{volume}{92}},
  \bibinfo{eid}{017902} (\bibinfo{year}{2004}).

\bibitem[{\citenamefont{Sanaka et~al.}(2006)\citenamefont{Sanaka, Resch, and
  Zeilinger}}]{sanaka:083601}
\bibinfo{author}{\bibfnamefont{K.}~\bibnamefont{Sanaka \emph{et al}}},
  \bibinfo{journal}{Phys. Rev. Lett.} \textbf{\bibinfo{volume}{96}},
  \bibinfo{eid}{083601} (\bibinfo{year}{2006}).

\bibitem[{\citenamefont{Resch et~al.}(2007)\citenamefont{Resch, O'Brien,
  Weinhold, Sanaka, Lanyon, Langford, and White}}]{resch:203602}
\bibinfo{author}{\bibfnamefont{K.~J.} \bibnamefont{Resch \emph{et al}}},
  \bibinfo{journal}{Phys. Rev. Lett.} \textbf{\bibinfo{volume}{98}},
  \bibinfo{eid}{203602} (\bibinfo{year}{2007}).

\bibitem[{\citenamefont{Peres}(1996)}]{PhysRevLett.77.1413}
\bibinfo{author}{\bibfnamefont{A.}~\bibnamefont{Peres}},
  \bibinfo{journal}{Phys. Rev. Lett.} \textbf{\bibinfo{volume}{77}},
  \bibinfo{pages}{1413} (\bibinfo{year}{1996}).

\bibitem[{\citenamefont{Slater}(2003)}]{slater-2003-5}
\bibinfo{author}{\bibfnamefont{P.~B.} \bibnamefont{Slater \emph{et al}}},
  \bibinfo{journal}{Phys. Rev. A} \textbf{\bibinfo{volume}{71}},
  \bibinfo{pages}{052319} (\bibinfo{year}{2005}).

\bibitem[{\citenamefont{Cabello et~al.}(2005)\citenamefont{Cabello, Feito, and
  Lamas-Linares}}]{cabello-2005-72}
\bibinfo{author}{\bibfnamefont{A.}~\bibnamefont{Cabello \emph{et al}}},
  \bibinfo{journal}{Phys. Rev. A} \textbf{\bibinfo{volume}{72}},
  \bibinfo{pages}{052112} (\bibinfo{year}{2005}).
 
\bibitem[{\citenamefont{Osenda and Raggio}(2005)}]{osenda-2005}
\bibinfo{author}{\bibfnamefont{O.}~\bibnamefont{Osenda}} \bibnamefont{and}
  \bibinfo{author}{\bibfnamefont{G.~A.} \bibnamefont{Raggio}},
  \bibinfo{journal}{Phys. Rev. A} \textbf{\bibinfo{volume}{72}},
  \bibinfo{pages}{064102} (\bibinfo{year}{2005}).

\bibitem[{\citenamefont{Jami and Sarbishei}(2006)}]{Jami:quant-ph0606039}
\bibinfo{author}{\bibfnamefont{S.}~\bibnamefont{Jami \emph{et al}}}
\eprint{quant-ph/0606039},
  (\bibinfo{year}{2006}).

\bibitem[{\citenamefont{Slater}(2007)}]{slater-2007}
\bibinfo{author}{\bibfnamefont{P.~B.} \bibnamefont{Slater}},
\eprint{quant-ph/0702134},
 (\bibinfo{year}{2007}).
 
 \bibitem[{\citenamefont{Ann and Jaeger}(2007)}]{ann-2007}
\bibinfo{author}{\bibfnamefont{K.}~\bibnamefont{Ann}} \bibnamefont{and}
  \bibinfo{author}{\bibfnamefont{G.}~\bibnamefont{Jaeger}},
\eprint{quant-ph/0707.4485},
  (\bibinfo{year}{2007}).

\bibitem[{\citenamefont{Hong et~al.}(1987)\citenamefont{Hong, Ou, and
  Mandel}}]{PhysRevLett.59.2044}
\bibinfo{author}{\bibfnamefont{C.~K.} \bibnamefont{Hong }},
  \bibinfo{author}{\bibfnamefont{Z.~Y.} \bibnamefont{Ou}}, \bibnamefont{and}
  \bibinfo{author}{\bibfnamefont{L.}~\bibnamefont{Mandel}},
  \bibinfo{journal}{Phys. Rev. Lett.} \textbf{\bibinfo{volume}{59}},
  \bibinfo{pages}{2044} (\bibinfo{year}{1987}).

\bibitem{Fockfilter} For our Fock-filter with reflectivity $R{=}r^2$, $\mathbf{P}_3(r){=}$
%
\begin{equation}
\begin{bmatrix} 
 r(2-3r^2) & 0 & 0 \\
0 &  r(1-2r^2) & 0 \\
0 & 0  & - r^3 \nonumber
\end{bmatrix}.
\end{equation}

\bibitem{waveplates} From Ref.~\cite{bogdanov:042303}, the waveplate action on a qutrit is:
\begin{equation}
\begin{bmatrix}
t^2			& \sqrt{2} t r		& r^2 \\
-\sqrt{2} t r^{*}	& |t|^{2}{-}|r|^{2}		& \sqrt{2} t^{*} r \\
r^{*2} 		& -\sqrt{2} t^{*} r^{*}	& t^{*2} 
\end{bmatrix}, \nonumber
\end{equation}
where $t{=}\cos \delta{+}i \sin \delta \cos 2 \alpha$, $r{=}i \sin \delta \sin 2 \theta$, and $\theta$ is the waveplate angle. For a half-wave plate, $\mathbf{H}_{3}(\theta)$, $\delta{=}\pi/2$; for a quarter-wave plate, $\mathbf{Q}_{3}(\theta)$, $\delta{=}\pi/4$.

\bibitem[{\citenamefont{Doherty and Gilchrist}(2007)}]{Gilchrist:2007lr}
\bibinfo{author}{\bibfnamefont{A.}~\bibnamefont{Doherty}} \bibnamefont{and}
  \bibinfo{author}{\bibfnamefont{A.}~\bibnamefont{Gilchrist}},
 \bibinfo{note}{in preparation},   (\bibinfo{year}{2007}).

\bibitem[{\citenamefont{O'Brien et~al.}(2004)\citenamefont{O'Brien, Pryde,
  Gilchrist, James, Langford, Ralph, and White}}]{obrien:080502}
\bibinfo{author}{\bibfnamefont{J.~L.} \bibnamefont{O'Brien \emph{et al}}},
  \bibinfo{journal}{Phys. Rev. Lett.} \textbf{\bibinfo{volume}{93}},
  \bibinfo{eid}{080502} (\bibinfo{year}{2004}).

\bibitem[{def()}]{definitions}
\bibinfo{howpublished}{Fidelity is
  $F(\rho,\sigma){\equiv}\{\mathrm{Tr}[\sqrt{\sqrt{\rho} \sigma
  \sqrt{\rho}}]\}^{2}$; linear entropy is $S_{L} {\equiv}$ $d
  (1{-}\mathrm{Tr}[\rho^2])/(d{-}1)$, where $d$ is the state dimension}.

\bibitem[{\citenamefont{White et~al.}(2007)\citenamefont{White, Gilchrist,
  Pryde, O'Brien, Bremner, and Langford}}]{measuringgates}
\bibinfo{author}{\bibfnamefont{A.~G.} \bibnamefont{White \emph{et al}}},
  \bibinfo{journal}{JOSA B} \textbf{\bibinfo{volume}{24}}, \bibinfo{pages}{172}
  (\bibinfo{year}{2007}).

\bibitem[{\citenamefont{Fulconis et~al.}(2007)\citenamefont{Fulconis, Alibart,h
  O'Brien, Wadsworth, and Rarity}}]{bristolpaper}
\bibinfo{author}{\bibfnamefont{J.}~\bibnamefont{Fulconis \emph{et al}}},
  \bibinfo{journal}{arXiv:quant-ph/0611232}
  (\bibinfo{year}{2007}).

\bibitem[{\citenamefont{et~al}(2007)}]{Weinhold}
\bibinfo{author}{\bibfnamefont{T.~J.~Weinhold \emph{et al},} }
 \bibinfo{note}{in preparation}, (\bibinfo{year}{2007}).

\bibitem[{neg()}]{negativity}
\bibinfo{howpublished}{Negativity is defined as $N{=}\emph{max}\{0,-\sum_i
  \lambda_i\}$ where $\lambda_i$ are the negative eigenvalues of the partial
  transpose of the target density matrix}.
  
  \bibitem[{\citenamefont{Knill et~al.}(2001)\citenamefont{Knill, Laflamme, and
  Milburn}}]{Knill:2001qy}
\bibinfo{author}{\bibfnamefont{E.}~\bibnamefont{Knill}},
\bibinfo{author}{\bibfnamefont{R.}~\bibnamefont{Laflamme}}, \bibnamefont{and}
 \bibinfo{author}{\bibfnamefont{G.~J.} \bibnamefont{Milburn}},
  \bibinfo{journal}{Nature} \textbf{\bibinfo{volume}{409}}
  (\bibinfo{year}{2001}). 
  
    \bibitem[{\citenamefont{Popescu}(2006)}]{popescu-2006}
\bibinfo{author}{\bibfnamefont{S.}~\bibnamefont{Popescu}},
  \eprint{quant-ph/0610043},
  (\bibinfo{year}{2006}).
  
  \bibitem[{\citenamefont{Bregman et~al.}(2007)\citenamefont{Bregman, Aharonov,
  Ben-Or, and Eisenberg}}]{bregman-2007}
\bibinfo{author}{\bibfnamefont{I.}~\bibnamefont{Bregman \emph{et al}}},
  \eprint{quant-ph/0709.3804},
  (\bibinfo{year}{2007}).
  
  \bibitem[{\citenamefont{Bishop and Byrd}(2007)}]{bishop2007}
\bibinfo{author}{\bibfnamefont{C.}~\bibnamefont{Bishop}} \bibnamefont{and}
  \bibinfo{author}{\bibfnamefont{M.~S.} \bibnamefont{Byrd}}
   \eprint{quant-ph/0709.0021},
  (\bibinfo{year}{2007}).
  
\bibitem{Ali07} M. Ali, et al., arxiv:0710.2238, (2007).
% A. R. P. Rau, Kedar Ranade, Disentanglement in qubit-qutrit systems

\end{thebibliography}
\end{document}